\documentstyle[epsfig]{aipproc}

\begin{document}
\title{New Models for X-Ray Synchrotron Radiation from the Remnant of Supernova 1006 AD}

\author{Kristy K. Dyer, Stephen P. Reynolds \& Kazimierz J. Borkowski}
\address{North Carolina State University\\
Physics Department, Raleigh NC 27695-8202\\}

\maketitle

\begin{abstract}
Galactic cosmic rays up to energies of around 10$^{15}$~eV are assumed to
originate in supernova remnants (SNRs). The shock wave of a young SNR like
SN 1006 AD can accelerate electrons to energies greater than 1 TeV, where
they can produce synchrotron radiation in the X-ray band.  A new model
({\it SRESC}) designed to model synchrotron X-rays from Type Ia supernovae can
constrain values for the magnetic-field strength and electron scattering
properties, with implications for the acceleration of the unseen
ions which dominate the cosmic-ray energetics. New
observations by ASCA, ROSAT, and RXTE have provided enormously improved data,
which now extend to higher X-ray energies. These data allow much firmer
constraints. We will describe model fits to these new data on SN 1006 AD,
emphasizing the physical constraints that can be placed on SNRs and on the
cosmic-ray acceleration process.
\end{abstract}

\section*{X-ray Synchrotron Emission from SNRs}

Young supernova remnants are in a period of transition to the Sedov-Taylor phase, sweeping up several times their original ejected mass as they expand into a circumstellar medium. The shocked gas at temperatures of 10$^7$ K emits a thermal spectrum (bremsstrahlung and lines) through interactions of electrons with ions. However, the synchrotron emission, which produces the radio spectrum, persists through the X-ray regime (and in some case dominates the thermal X-rays \cite{koyama95}). 

In order to study the energetics of the accelerated electrons and protons in the shock, as well as to obtain accurate abundances from thermal models, it is imperative to understand this synchrotron spectrum. SNR observations show that the synchrotron emission drops {\it significantly} below a powerlaw at X-ray frequencies, implying a curved synchrotron spectrum. To accurately describe the synchrotron spectrum we have developed the following models:

\begin{itemize}
\item {\bf SRCUT} A homogeneous population of electrons, with an exponentially
cut off power-law energy distribution, radiating in a constant
magnetic field, produces the sharpest physically plausible cutoff in
the emitted spectrum.  Given a radio flux and spectral index, the
models depend on a single parameter: the frequency at which the
spectrum has dropped by a factor of 10 below the power-law
extrapolation.  If a remnant emits primarily thermal X-rays, {\it SRCUT} will give the
maximum nonthermal-electron energies allowed, so as not to exceed
observed X-ray fluxes.  {\it SRCUT} fits to 14 galactic SNRs give electron
upper limits of 100 TeV or less, well below the 1000 TeV ``knee'' in the cosmic-ray (ion) spectrum.\cite{keohane99}

\item {\bf SRESC} In some remnants the energy of the most energetic electrons will be limited because the shock cannot effectively scatter particles above a certain gyroradius. The {\it SRESC} model describes Sedov expansion into a uniform magnetic field, appropriate for Type Ia and late core-collapse SNR. Details are discussed in Reynolds (1998).\cite{reynolds98}

\end{itemize}

These models are currently available and will be distributed in the next release of XSPEC. The XSPEC model formats should also be compatible with CXC, the Chandra software. More sophisticated models, as described in Reynolds (1998)\cite{reynolds98}, will be released in the future.

\section*{The Escape Model}

The electron scattering required for acceleration to high enough energies to produce synchrotron radiation may become less efficient above some energy. Or, in the limiting case, electrons above some energy may freely escape the SNR. In the escape model we assume that magnetohydrodynamic scattering waves are much weaker above some wavelength $\lambda_{max}$. Since electrons with gyroradius $r_g$ scatter resonantly with waves of wavelengths $\lambda = 2\pi r_g$, electrons will escape once their energy reaches $E_{max}$ given by:

\begin{equation}
E_{max} = \lambda_{max}B_1/4
\end{equation}

\noindent where $B_1$ is the upstream magnetic field strength.
\smallskip

Reynolds (1998) produced a detailed model for SNR emission with electron energies limited by escape, which includes correct accounting for variation of shock-acceleration efficiency and post-shock radiative and adiabatic losses, assuming Sedov dynamics.

The model has three parameters: 1) the radio flux measurement at 1 GHz, 2) $\alpha$, the radio spectral index (flux density $\propto \nu^{-\alpha}$), 3) a characteristic rolloff frequency, the frequency at which the spectrum has dropped by approximately 10 below a straight powerlaw. The spectral index and 1 GHz flux for SNR are fixed by observations. For Galactic SNRs, they can be found at Green's website.\cite{green98}

\section*{SN 1006 AD}

\begin{figure}[b!] 
\centerline{\epsfig{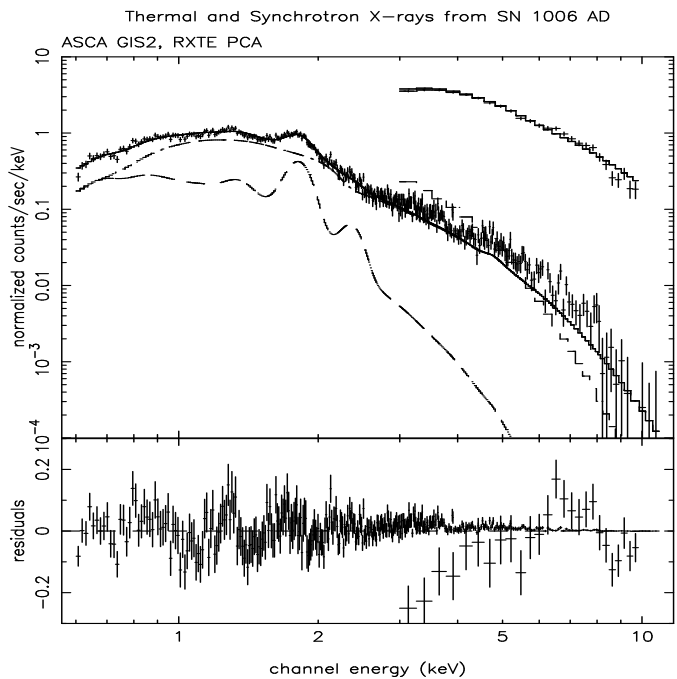}}
\vspace{10pt}
\caption{X-ray observations of SN 1006 by RXTE PCA (upper) and ASCA GIS2 (lower) fit with {\it SRESC} and {\it VSHOCK}. The separate contributions of the {\it VSHOCK} and {\it SRESC} models are shown (broken lines) below the sum (solid lines).}
\label{fig1}
\end{figure}

\begin{figure}[b!] 
\centerline{\epsfig{file=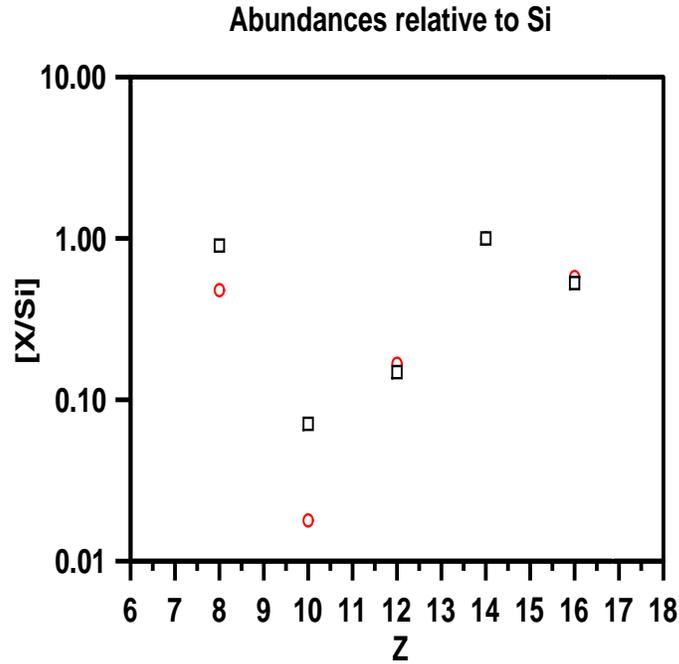,height=3.5in,width=3.5in}}
\vspace{10pt}
\caption{Abundances (by mass, normalized to silicon) from a {\it VSHOCK} + {\it SRESC} fit (circles) compared to theoretical predictions (squares) from Nomoto, Thielemann \& Yokoi (1984) for a Type I supernova model (W7).}
\label{fig2}
\end{figure}

SN 1006 AD is an example of a young remnant where the synchrotron emission dominates in the X-ray regime\cite{koyama95}. Early observations, with low signal to noise, could be adequately fit by curved synchrotron models \cite{reynolds96}. Observations from RXTE are a 10-fold improvement in sensitivity and provide a more sensitive test. 

ASCA and RXTE observations of SN 1006 AD can be adequately fit with two components: the escape model and a thermal plane-parallel nonequilibrium-ionization shock model ({\it VSHOCK}) with variable abundances (Borkowski et al., in preparation). The model fit depends relatively weakly on the flux at 1 GHz. For SN 1006 AD the flux was fixed at the observed value of 19 Jy\cite{green98}. The column density was fixed at $5\times10^{20}$ cm$^{-2}$, a value consistent with optical and ROSAT PSPC observations. The best fit, shown in Figure 1, has a $\alpha$=0.58 and the rolloff frequency of 1.7$\times 10^{17}$ Hz with a $\chi^2$ of 665 for 376 degrees of freedom. The {\it VSHOCK} obtains abundances (relative to solar) of O 0.27, Ne 0.16, Mg 1.56, Si 6.43 and S 8.16, compared in Figure 2 to predictions for Type Ia supernova by Nomoto et al.(1984).\cite{nomoto84} From the synchrotron formula, $\nu_m \propto E^2B$, and 
equation 1, assuming a magnetic field of 3$\mu$G, we calculate a $\lambda_{max}$ of 1.2$\times 10^{17}$ cm.

X-ray emission from SN1006 AD is well fit by synchrotron models with electron energies limited by escape. Comparing the escape model to future models that are limited by radiation losses or SNR age will give us a better understanding of synchrotron emission at X-ray energies. Our future work involves applying these models to SNRs and making them widely available to the community by distribution through XSPEC.

\medskip

{\it Thanks to G. Allen for sharing RXTE X-ray observations of SN1006 AD and J. Keohane for research notes and advice.  This research is supported by NASA grant NAG5-7153 and NGT5-65 through the Graduate Student Researchers Program. }

\end{document}